\documentclass[final,3p,times,twocolumn]{elsarticle}

\usepackage{graphicx}
\usepackage{float}
\newlength\figurewidth
\usepackage{caption}
\usepackage{subcaption}

\usepackage{pgfplots}
\usepgfplotslibrary{units}
\usepgfplotslibrary{polar}
\pgfplotsset{compat=1.14}

\usepackage{tikz}
\usetikzlibrary{arrows}
\usetikzlibrary{arrows.meta}
\usetikzlibrary{calc}
\usetikzlibrary{shapes}
\usetikzlibrary{positioning}
\usetikzlibrary{spy}
\usetikzlibrary{decorations.pathmorphing}

\usepackage{amsmath}
\usepackage{amssymb}
\usepackage{mathrsfs} 
\renewcommand{\H}{$\mathscr H$}
\usepackage{siunitx}
\usepackage[pdftex, pagebackref=false, bookmarks=true, bookmarksnumbered=false, pdfstartview = FitH, plainpages=false, colorlinks=false, pdfborder={0 0 0}]{hyperref}
\usepackage{xcolor}

\definecolor{niceblue}{RGB}{55,126,184}
\definecolor{nicered}{RGB}{228,26,28}
\definecolor{nicegreen}{RGB}{77,175,74}
\definecolor{niceorange}{RGB}{255,127,0}
\definecolor{nicepurple}{RGB}{152,78,163}
\definecolor{niceyellow}{RGB}{255,255,51}

\usepackage[plain]{algorithm}
\usepackage{algpseudocode}
\algnewcommand{\LeftComment}[1]{\State \(\triangleright\) #1}
\algblock{ParFor}{EndParFor}
\algblock{ParForAll}{EndParFor}
\algnewcommand\algorithmicparfor{\textbf{parfor}}
\algnewcommand\algorithmicparforall{\textbf{parfor all}}
\algnewcommand\algorithmicpardo{\textbf{do}}
\algnewcommand\algorithmicendparfor{\textbf{end\ parfor}}
\algrenewtext{ParFor}[1]{\algorithmicparfor\ #1\ \algorithmicpardo}
\algrenewtext{ParForAll}[1]{\algorithmicparforall\ #1\ \algorithmicpardo}
\algrenewtext{EndParFor}{\algorithmicendparfor}

\usepackage{xargs}
\usepackage[colorinlistoftodos,prependcaption,textsize=footnotesize]{todonotes}
\newcommandx{\unsure}[2][1=]{\todo[linecolor=red,backgroundcolor=red!25,bordercolor=red,#1]{#2}}

\usepackage{lineno}

\journal{Engineering Analysis with Boundary Elements}

\begin{document}

\begin{frontmatter}


\title{Simple and efficient GPU parallelization of existing \H-Matrix accelerated BEM code}



\author[address1]{Kerstin Vater\fnref{address2}}
\ead{kerstin.vater@tuhh.de}
\author[address2]{Timo Betcke}
\ead{t.betcke@ucl.ac.uk}
\author[address3]{Boris Dilba}
\ead{dilba@novicos.de}
\address[address1]{Hamburg University of Technology, Dynamics Group, Schlo{\ss}m{\"u}hlendamm 30, 21073 Hamburg, Germany}
\address[address2]{Department of Mathematics, University College London, Gower Street, London WC1E  6BT, UK}
\address[address3]{Novicos GmbH, Kasernenstra{\ss}e 12, 21073 Hamburg}

\begin{abstract}
In this paper, we demonstrate how GPU-accelerated BEM routines can be used in a simple black-box fashion to accelerate fast boundary element formulations based on Hierarchical Matrices (\H-Matrices) with ACA (Adaptive Cross Approximation). In particular, we focus on the expensive evaluation of the discrete weak form of boundary operators associated with the Laplace and the Helmholtz equation in three space dimensions.
The method is based on offloading the CPU assembly of elements during the ACA assembly onto a GPU device and to use threading strategies across ACA blocks to create sufficient workload for the GPU. The proposed GPU strategy is designed such that it can be implemented in existing code with minimal changes to the surrounding application structure. This is in particular interesting for existing legacy code that is not from the ground-up designed with GPU computing in mind.

Our benchmark study gives realistic impressions of the benefits of GPU-accelerated BEM simulations by using state-of-the-art multi-threaded computations on modern high-performance CPUs as a reference, rather than drawing synthetic comparisons with single-threaded codes. Speed-up plots illustrate that performance gains up to a factor of $\num{5.5}$ could be realized with GPU computing under these conditions. This refers to a boundary element model with about $\num{4}$ million unknowns, whose \H-Matrix weak form associated with a real-valued (Laplace) boundary operator is set up in only $\num{100}$ minutes harnessing the two GPUs instead of $\num{9}$ hours when using the 20 CPU cores at disposal only. The benchmark study is followed by a particularly demanding real-life application, where we compute the scattered high-frequency sound field of a submarine to demonstrate the increase in overall application performance from moving to a GPU-based ACA assembly.
\end{abstract}


\end{frontmatter}


\section{Introduction}
\label{sec:intro}

Many problems in engineering science referring to an equilibrium state in a homogeneous medium can be modeled by the Boundary Element Method (BEM). The fundamental idea is to express the solution only in terms of values on the boundary of the calculation domain. This feature makes the BEM especially suitable for unbounded domains, as they are frequently encountered in acoustics, electrostatics and fluid dynamics.

In the course of a boundary element simulation, a system of linear equations is set up to find the unknown part of the boundary data. For this purpose, it is necessary to evaluate the discrete weak formulations of the boundary operators involved. Their explicit computation is usually the most expensive part of a boundary element simulation, since classically $\mathcal{O}\left(N^2\right)$ integrals have to be evaluated numerically for a BEM problem with $N$ elements. Hierarchical Matrices (\H-Matrices) based on the Adaptive Cross Approximation (ACA) approach can reduce this complexity to $\mathcal{O}\left(N\log N\right)$. Together with the relative simple implementation of ACA it has therefore become the method of choice for many large-scale industrial applications.

With the rise of Graphics Processing Units (GPUs) to be used for scientific computations, entirely new possibilities have opened up to further accelerate boundary element simulations. Graphics cards have been originally designed for image rendering purposes, where tremendous amounts of light-weight, independent tasks are processed in parallel. It turns out that numerical integration routines in the BEM operate in a very similar way, in the sense that they are also massively invoked and execute in a Single Instruction Multiple Data (SIMD) fashion. This circumstance suggests that GPU computing may benefit the assembly procedure of the discretized weak form of boundary operators in the BEM, and therefore further speed up the expensive setup of the equation system.

The acceleration of boundary element simulations has been a subject of research for decades. Until about 2010, activities in this area were mainly focused on the development of fast approximation algorithms. Popular approaches such as the Fast Multipole Method (FMM) \cite{greengard1987fast, liu2009fast, keuchel2017hp} or \H-Matrices \cite{borm2003introduction, hackbusch1999sparse} have proven capable of significantly reducing the computational effort associated with boundary element calculations.

One of the first attempts to speed up the BEM using graphics hardware has been made in 2009 by \citet{takahashi2009gpu} when they presented a GPU-accelerated BEM formulation based on the GPU programming platform CUDA to solve the Helmholtz equation in three dimensions. Two years later, \citet{labaki2011constant, labaki2010bem} published a review on three distinct GPU implementations addressing two-dimensional potential and electrostatic problems.

In 2011, \citet{yokota2011biomolecular} employed the BEM to investigate biomolecular electrostatic interactions governed by a Poisson equation. For this purpose, they designed a solver using GPU hardware acceleration on top of an FMM code. Computational experiments with billion-scale problems demonstrated good parallel efficiency on up to $\num{512}$ GPUs. Shortly after, the authors implemented an auto-tuning mechanism \cite{yokota2012scaling}, which enabled reasonable scaling of their FMM code on up to $\num{4000}$ GPUs in the context of particle-based turbulence simulations.

For the GPU acceleration of \H-Matrix compression Christophersen proposed in 2012 an OpenCL-based algorithm that used interpolation techniques for the far-field and not ACA. Recently, \citet{boerm2015approximation} suggested a hybrid adaptive algorithm based on a technique called Green's cross approximation to achieve very good GPU performance. However, this technique makes a redesign of the \H-Matrix code necessary.

The main objective of this paper is to reveal how existing \H-Matrix BEM codes can be upgraded with GPU computing to accelerate the dense and \H-Matrix weak-form assembly of all four commonly encountered boundary operators associated with the three-dimensional Laplace and Helmholtz equations. These are the single- (SLP) and double-layer potential (DLP) operators, as well as the adjoint double-layer potential (ADLP) and hypersingular (HYPS) boundary operators. We will show that it is actually reasonable to apply GPU computing within the flexible framework of an existing BEM library framework running on a desktop computer with minimum code changes, and give realistic impressions of what can be expected from this approach.

For this purpose, the GPU programming interface CUDA C/C++ will be employed, which is included in the CUDA Toolkit \cite{cudaToolkit} version 8.0. The implementation is based on an experimental branch of the Galerkin boundary element library Bempp \cite{smigaj2015solving} version 3.1 (\url{www.bempp.com}), which provides flexible solutions for various BEM applications in a black-box manner. Therefore, the possibility to choose boundary elements of variable polynomial order will be preserved on the GPU. The system under consideration is a workstation equipped with two Intel Xeon processors E5-2670 v2 \cite{intelXeonProcessorSpecs} with ten physical cores each, operating at $\SI{2.5}{\giga\hertz}$ processor base frequency, as well as two NVIDIA GeForce GTX TITAN Black boards \cite{nvidiaTitanGPUSpecs} with a full Kepler GK110 GPU implementation each, and about $\SI{200}{\giga\byte}$ of system memory.

The remainder of this paper is structured as follows. Section \ref{sec:bempp} briefly reviews the mathematical framework of \H-Matrix-based Galerkin boundary element formulations. Sections \ref{sec:denseImpl} and \ref{sec:hmatImpl} discuss how the relevant parts of the underlying C++ code can be adapted such that the dense and \H-Matrix-based assembly routines run on GPUs. Benchmark results in Section \ref{sec:benchmark} illustrate the performance of the proposed algorithms for various Laplace and Helmholtz boundary operators, where both the complete and the \H-Matrix weak-form construction are investigated.
We conclude the benchmarks with a realistic industry example to compute the scattered sound field of a submarine. In Section \ref{sec:conclusions}, we present a summary and an outlook on ongoing work.

\section{Galerkin-based boundary integral operators and their discretization}
\label{sec:bempp}

In this work, we consider Galerkin discretizations of the single-layer potential (SLP), double-layer potential (DLP), adjoint double-layer potential (ADLP), and hypersingular (HYPS) boundary operators for three-dimensional Laplace and Helmholtz problems. The corresponding discrete matrices are defined as follows.
\begin{align}
	\left[S_h\right]_{i,j}
    &= \int_{\Gamma_j} \int_{\Gamma_i} \psi_i\left(x\right) g\left(x,y\right) \phi_j\left(y\right) \mathrm{d}s_y \mathrm{d}s_x \,,
    \label{eq:slp}\\
	\left[K_h\right]_{i,j}
    &= \int_{\Gamma_j} \int_{\Gamma_i} \psi_i\left(x\right) \frac{\partial g\left(x,y\right)}{\partial n_y} \phi_j\left(y\right) \mathrm{d}s_y \mathrm{d}s_x \,,
    \label{eq:dlp}\\
	\left[K'_h\right]_{i,j}
    &= \int_{\Gamma_j} \int_{\Gamma_i} \psi_i\left(x\right) \frac{\partial g\left(x,y\right)}{\partial n_x} \phi_j\left(y\right) \mathrm{d}s_y \mathrm{d}s_x \,.
    \label{eq:adlp}\\
	\left[D_{h}\right]_{i,j}
    &= \int_{\Gamma_j} \int_{\Gamma_i} g\left(x,y\right) \left[\mathrm{curl}_{\Gamma}\psi_i\left(x\right) \cdot \mathrm{curl}_{\Gamma}\phi_j\left(y\right)\right. \nonumber\\
    &\left.- k^2 \psi_i\left(x\right)n_x \cdot \phi_j\left(y\right)n_y\right] \mathrm{d}s_y \mathrm{d}s_x \,.
    \label{eq:hypsHelmholtz}
\end{align}
In the above equations, $n$ denotes the unit normal vector at a specified point on the boundary $\Gamma$ of the solution domain, and $\phi_i$ and $\psi_j$ represent real basis functions of the domain and test space, respectively. We note that for the hypersingular operator the functions $\phi_j$ and $\psi_i$ must be at least continuous and piecewise linear. The above definition for $D$ can then be obtained from the normal derivative of the double-layer potential, followed by integration by parts.

The Green's function of the Helmholtz equation ($-\Delta u - k^2 u = 0$) in 3D is defined as
\begin{align}
	g\left(x,y\right) = \frac{\exp\left(\mathrm{i}k\left|x-y\right|\right)}{4\pi\left|x-y\right|} \,.
\end{align}
The Laplace equation in 3D ($-\Delta u = 0$) has the same Green's function with the wavenumber $k$ set to zero.


\subsection{Hierarchical Matrices}
\label{subsec:bemppHmat}

\begin{figure}[H]
  \centering
  \includegraphics[width=8cm]{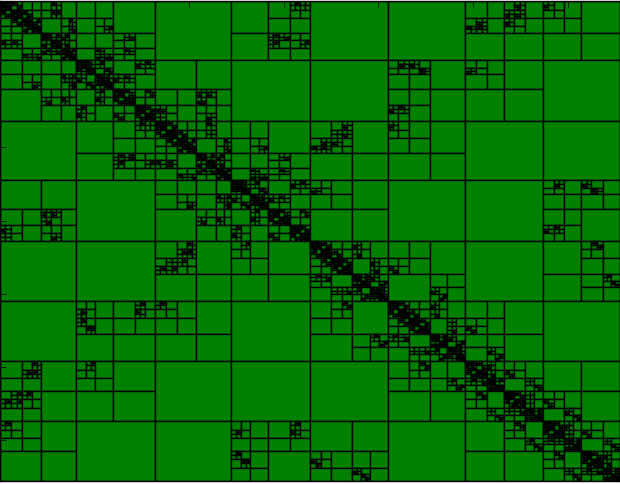}
  \caption{Hierarchical partition of a matrix depending on a geometric admissibility condition.}
\label{fig:hierarchicalPartition}
\end{figure}

In the following, we briefly review ACA-based \H-Matrix assembly \cite{bebendorf2008hierarchical} of boundary operators. The basic idea is to approximate the fully-populated matrices \eqref{eq:slp} to \eqref{eq:hypsHelmholtz} by using only a few of the matrix entries. For this purpose, a tree-based geometric partitioning is introduced. Leafs of the tree correspond to matrix blocks, which are either admissible with respect to a distance and block diameter dependent admissibility condition, or inadmissible. Admissible blocks can be compressed. Inadmissible blocks are stored in dense mode. An example partitioning is shown in Figure \ref{fig:hierarchicalPartition}. Large blocks are admissible while most of the very small blocks are inadmissible.


Several different approaches exist to treat the admissible blocks. The probably most popular one is known as the Adaptive Cross Approximation (ACA) procedure. An introduction and analysis of the ACA algorithm is given in \citet[Section~3.4]{bebendorf2008hierarchical}. In the following, we briefly recap the principle of ACA as it will be important to understanding the GPU algorithm later on.

Consider an admissible leaf block $A\in\mathbb{R}^{m\times n}$. Starting from $R_0:=A$, the recursion formula
\begin{align}
  R_{k+1} := R_k - \left[\left(R_k\right)_{i_k,j_k}\right]^{-1} \left(R_k\right)_{1:m,j_k} \left(R_k\right)_{i_k,1:n}
  \label{eq:acaRecursionFormula}
\end{align}
yields a low-rank representation of the block $A$. The index ranges $\left(R_k\right)_{i,1:n}$ and $\left(R_k\right)_{1:m,j}$ denote the $i$-th row and the $j$-th column of $R_k$, respectively. Essentially, this means that we have to find a non-zero pivot $\left(i_k,j_k\right)$ in $R_k$, which is used as a scaling factor for an outer product of the $i_k$-th row and the $j_k$-th column. The result is then subtracted from $R_k$ to get $R_{k+1}$, and so forth.

The column index $j_k$ is chosen by finding the maximum element of the current row $i_k$ according to
\begin{align}
  \left|\left(R_{k-1}\right)_{i_k,j_k}\right| = \max_{j=1,\ldots,n} \left|\left(R_{k-1}\right)_{i_k,j}\right| \,.
\end{align}
The selection of the row index $i_k$, however, turns out to be more complicated. It is described in detail by \citet[Section~3.4.3]{bebendorf2008hierarchical}.

On closer inspection of the recursion \eqref{eq:acaRecursionFormula} it becomes apparent that the $k$-th step only incorporates matrix entries in the $j_k$-th column and the $i_k$-th row to compute a better approximation $R_{k+1}$. Thus, it is not necessary to determine the full matrix $R_k$ in each step, except from a few of the original entries of $A$. Taking advantage of this property, the ACA algorithm takes the following form.
\begin{algorithm}[H]
\hrulefill
\vspace{-4pt}
\begin{algorithmic}
  \Procedure{ACA~}{admissible leaf block $A\in\mathbb{R}^{m\times n}$}
  \State Initialization $k \gets \num{1}$, $Z \gets \emptyset$
  \Repeat
    \State Determine row index $i_k$
    \State Compute selected row $\tilde{v}_k \gets a_{i_k,1:n}$
    \LeftComment Subtract scaled preceding results
    \For{$l=1,\ldots,k-1$}
      \State $\tilde{v}_k \gets \tilde{v}_k - \left(u_l\right)_{i_k} v_l$
    \EndFor
    \State Add $i_k$ to $Z$
    \If{$\tilde{v}_k$ does not vanish}
      \State Find column index $j_k \gets \max_{j=1,\ldots,n}
      \left|\left(\tilde{v}_k\right)_j\right|$
      \State $v_k \gets \left(\tilde{v}_k\right)^{-1}_{j_k} \tilde{v}_k$
      \State Compute selected column $u_k \gets a_{1:m,j_k}$
      \LeftComment Subtract scaled preceding results
      \For{$l=1,\ldots,k-1$}
        \State $u_k \gets u_k - \left(v_l\right)_{j_k} u_l$
      \EndFor
      \State $k \gets k+1$
    \EndIf
  \Until{stopping criterion is met \\ \hspace{1.25cm} or $Z$ contains all $m$ rows}
\EndProcedure
\end{algorithmic}
\vspace{-8pt}
\hrulefill
\end{algorithm}
Here, $u_k$ denotes the $j_k$-th column of $R_{k-1}$, i.e. $\left(R_{k-1}\right)_{1:m,j_k}$, and $\tilde{v}_k$ is the $i_k$-th row of $R_{k-1}$, i.e. $\left[\left(R_{k-1}\right)_{i_k,1:n}\right]^T$. The matrix $S_k := \sum_{l=1}^k u_l v_l^T$ eventually represents an approximation of $A = S_k + R_k$.

The rank of $S_k$ is bounded by the number of updates $k$. The rank also determines the accuracy of the approximation. Therefore, a maximum number of iterations $k_{\mathrm{max}}$ can be defined adaptively as a stop criterion, such that a predefined level of accuracy is ensured. However, the algorithm usually converges already after a few steps.

Note that for practical application the prototype ACA algorithm as outlined above is extended by several features, such as heuristic sub-block detection and zero-block identification.
\section{Dense GPU parallelization}
\label{sec:denseImpl}

In the following section, we turn to the major concern of this paper, the development of simple and effective GPU-accelerated weak-form assembly routines. First, we will give a very brief introduction into GPU computing with CUDA. We then describe the host program managing the deployment of accelerator hardware, as well as the actual GPU integration routines. We begin with the boundary operators as introduced in Section \ref{sec:bempp}, where the discrete weak formulation is represented by a fully-populated matrix. The resulting code parts will serve as the basis for the following \H-Matrix weak-form assembly.

\subsection{GPU programming with CUDA}
\label{sec:cuda}

The term ``CUDA'' originally stood for ``Compute Unified Device Architecture'', and is essentially a parallel computing platform with a dedicated application programming interface (API) developed by NVIDIA since the late 1990s. The programming model enables scientists and engineers to access graphics processors for GPGPU purposes in a straightforward way. The CUDA API is built on top of the high-level programming language C/C++, which is commonly used for scientific software development and High Performance Computing purposes.

The general concept of GPU programming with CUDA is that of a \textit{host} program executing on the CPU that manages computational tasks performed on \textit{devices}. Therefore, the terms CPU/host and GPU/device are used interchangeably in the following.

In order to get work done by the GPU, the host invokes a special function called \textit{kernel}. A CUDA kernel is characterized by the fact, that it is called from the host program and executes on the device. Each parallel thread on the GPU then launches an instance of the kernel, where work is spread across the requested compute resources via process indices. Detailed CUDA programming guides are freely available from the NVIDIA website \cite{cudaProgrammingGuide, cudaBestPracticesGuide}, and will therefore not be explicated here.

\subsubsection{Hardware architecture of GPUs}
For code development we have used TITAN Black graphics cards which are based on the NVIDIA Kepler GK110 microarchitecture \cite{nvidiaKeplerArch}, which hosts 15 Streaming Multiprocessor Units (SMX). Each SMX features 192 single precision CUDA cores, as well as 64 double precision units. This ``precision gap'' is the reason, why calculations in single precision typically run much faster on GPUs than double precision computations. The amount of fast accessible on-chip memory is restricted to 48 KB assigned to the L1 cache, 8 KB read-only data cache, and 1536 KB L2 cache memory. Note that the cache of a GPU multiprocessor is generally very small compared to modern CPU configurations. For comparison, one core of the Intel Xeon processors E5-2670 v2 has 25 MB cache at its disposal. This fact makes an efficient memory management on the GPU essential for good performance.

\subsubsection{Thread parallelism in CUDA}
In CUDA, parallel tasks are mapped to grids of blocks with blocks of threads. The described multiprocessors schedule tasks in \textit{warps} of 32 threads each, which execute in a SIMD fashion. Hence, branching in individual threads inside a warp (branch divergence) leads to slow down as CUDA runs through both branches in an if/then/else statement and discards results as necessary. 
Thus, GPU computing is made for identical, parallelizable tasks.

\subsection{Host code}
\label{subsec:denseImplHost}

Each element of the global BEM matrix consists of sums of integrals over test and trial elements. 
The numerical evaluation of these raw weak-form integrals is clearly the most expensive part of the matrix assembly procedure. Therefore, we source out this subroutine to the GPU, while the subsequent summation of element contributions to the respective degrees of freedom in the matrix is left to the CPU.

\subsubsection{Basic procedure}
The host code of the GPU-accelerated dense-matrix weak-form assembly can be outlined as follows. First, the total amount of work associated with all pairs of test and trial elements is distributed evenly over the available devices. This approach assumes that the GPUs which are attached to the system are all of the same type, which is usually the case for modern high-performance workstations and computer clusters.

The participating devices are covered by a parallel loop, where each GPU runs through the following steps. First, the respective device is initialized. This procedure includes the transfer of the raw grid data with vertex coordinates and element definitions to the global device memory space. Since in BEM we deal with surface grids which are typically not very large in size we always transfer the complete grid information to a device. Geometry information such as normal vectors and Jacobians are computed on the device and stored as 1D structures
according to the principle of coalesced access to global device memory \cite[Section~5.3]{cudaProgrammingGuide}.

Moreover, the relevant basis functions (and if necessary also their spatial derivatives) are evaluated at the quadrature points defined with respect to the unit triangle. These values reside in the global device memory space, as well.

Numerical quadrature weights, however, are stored in constant memory. These are only a few values, which are accessed by all GPU threads at the same time. Hence, this type of memory suggests itself. One practical drawback of constant memory is that it needs to be defined as a static array, the size of which must be known at compile time. Therefore, we impose a limit of six double-precision floating-point numbers in the first place, which corresponds to a quadrature order of $4$ on plane triangular boundary elements using a symmetric Gauss triangle quadrature rule.

After the initialization process is completed, each device subdivides its work package further into chunks of element pairs. In this way, the evaluation of unassembled matrix entries on the GPU can be overlapped with the host-side global assembly of elementary contributions to keep the available hardware busy and decrease the total execution time.

The unassembled GPU results related to a chunk are stored in \textit{zero-copy} host memory. This type of memory is advantageous in this case, because only two result buffers of the size corresponding to one chunk need to be allocated in the beginning of the process. Furthermore, integral values are written back only once at the end of the CUDA kernel in a coalesced fashion. Thus, zero-copy memory is supposed to be the fastest approach.

The chunks of element pairs are now processed successively. In doing so, the numerical integration and the global assembly procedure are considered as two distinct tasks, where the former is basically handled by the GPU and the latter is assigned to the CPU. These two tasks are part of a parallel TBB \cite{intelTbb} task group and can execute simultaneously. This means that one chunk passes through the global assembly process on the CPU, while the next bunch of element pairs is already integrated on the device. This approach helps to keep all available hardware as busy as possible.

Nevertheless, it turns out that either task necessarily limits the overall performance. Obviously, results can be assembled only after the computation of the corresponding values has been completed. This is ensured through a barrier between the integration step and the assembly step associated with the same chunk.

\begin{figure}[t]
  \centering
  \tikzset{%
  >={Latex[width=2mm,length=2mm]},
            line/.style = {draw, ->},
            base/.style = {rectangle, rounded corners, draw=black,
                           minimum width=2.5cm, minimum height=0.8cm,
                           text centered, font=\sffamily, anchor=north},
       startstop/.style = {base, fill=red!20},
       integrate/.style = {base, minimum width=2.2cm, fill=white},
        assemble/.style = {base, minimum width=2.2cm, fill=white},
         process/.style = {base, fill=orange!15}
}

\begin{tikzpicture}[node distance=1.25cm,
    every node/.style={fill=white, font=\sffamily}, align=center, anchor=north]
    
  \node (gpu) [rectangle, rounded corners, draw=black, fill=green!15,
  minimum width=2.6cm, minimum height=5.5cm, anchor=north]
  at (-1.5cm, -5.5cm) {}; 
  \node[anchor=north west, fill=none] at (gpu.north west) {GPU};

  \node (cpu) [rectangle, rounded corners, draw=black, fill=cyan!20,
  minimum width=2.6cm, minimum height=5.5cm, anchor=north]
  at ( 1.5cm, -5.5cm) {}; 
  \node[anchor=north west, fill=none] at (cpu.north west) {CPU};
  
  \node (start)             [startstop]     {\footnotesize Assembly starts};
  \node (dist)              [below of=start, text width=5cm, minimum width=3cm,
                             yshift=0.25cm]		
    {\footnotesize Distribute integrals equally over all available devices};
  \node (device1Block)      [process, below of=start, yshift=-1cm]
                                        {\footnotesize Devices 1, 2, ...};
  \node (init1Block)        [process, below of=device1Block, font=\ttfamily]
                                               {\footnotesize initialize()};
  \node (chunk1)            [below of=init1Block, text width=5cm, yshift=0.25cm]
                {\footnotesize Process integrals in chunks $\left(i\right)$}; 
  \node (dev1int0Block)     [integrate, below of=chunk1, yshift=-0.5cm, 
                             xshift=-1.5cm, font=\ttfamily]
                                               {\footnotesize integrate(0)};
  \node (dev1int1Block)     [integrate, below of=dev1int0Block, font=\ttfamily]
                                               {\footnotesize integrate(1)};
  \node (dev1int2Block)     [integrate, below of=dev1int1Block, font=\ttfamily]
                                               {\footnotesize integrate(2)};
  \node (dev1assem0Block)   [assemble, right of=dev1int1Block, yshift=0cm,
                             xshift=1.75cm, font=\ttfamily]
                                                {\footnotesize assemble(0)};
  \node (dev1assem1Block)   [assemble, right of=dev1int2Block, xshift=1.75cm,
                             font=\ttfamily]    {\footnotesize assemble(1)};
  \node (dev1assem2Block)   [assemble, below of=dev1assem1Block, font=\ttfamily]
                                                {\footnotesize assemble(2)};
  \node (ActivityDestroyed) [startstop, below of=start, yshift=-10cm]
                                          {\footnotesize Assembly finished};
                                          
  \draw[-]              (start) -- (dist);
  \draw[->]              (dist) -- (device1Block);
  \draw[->]      (device1Block) -- (init1Block);
  \draw[-]         (init1Block) -- (chunk1);
  \draw[->]      (chunk1.south) ++ (-1.5cm, 0) --(dev1int0Block);
  \draw[->]     (dev1int0Block) -- (dev1int1Block);
  \draw[->]     (dev1int1Block) -- (dev1int2Block);
  \draw[->]     (dev1int2Block) |- (ActivityDestroyed);
  \draw[->]      (chunk1.south) ++ (1.5cm, 0) -- (dev1assem0Block);
  \draw[->]   (dev1assem0Block) -- (dev1assem1Block);
  \draw[->]   (dev1assem1Block) -- (dev1assem2Block);
  \draw[->]   (dev1assem2Block) |- (ActivityDestroyed);
\end{tikzpicture}
  \caption{Dense weak-form assembly procedure}
  \label{fig:denseAssemblyProcedure}
\end{figure}
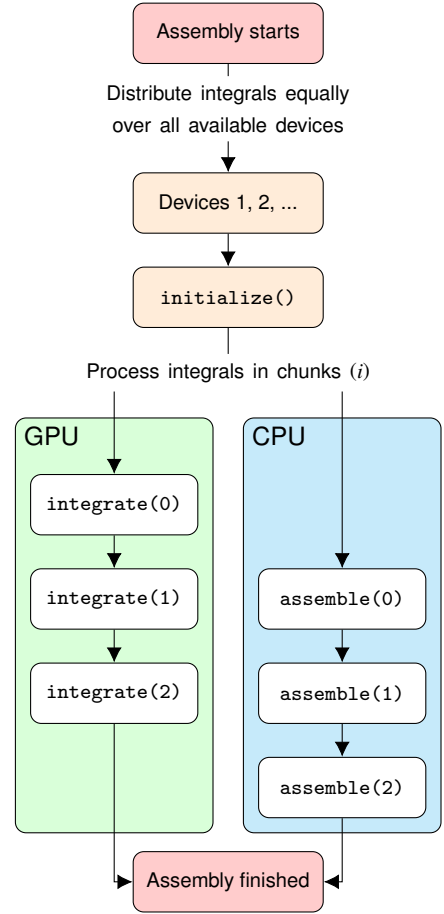

The described procedure is summarized in Figure \ref{fig:denseAssemblyProcedure}, as well as in the pseudo code below:
\begin{algorithm}[H]
\hrulefill
\vspace{-4pt}
\begin{algorithmic}
  \Procedure{Dense Assembly}{}
  \State Allocate result matrix $A$
  \State Distribute integrals over devices
  \ParForAll{devices}
    \State Initialize device
    \State Set up parallel task group
    \ForAll{chunks of integrals $b$} \textbf{simult.}
      \LeftComment Integrate $b_i$ on device and wait
      \ParForAll{integrals $r_j$ in $b_{i}$}
        \State $r_j \gets$ GPU results
      \EndParFor
      \LeftComment Assemble $b_{i-1}$ into matrix $A$ on host
      \ParForAll{integrals $r_j$ in $b_{i-1}$}
        \If{$r_j$ is singular}
          \State $A \gets \tilde{r}_j$ (CPU results) 
        \ElsIf{integral is regular}
          \State $A \gets r_j$ (GPU results)
        \EndIf
      \EndParFor
    \EndFor
    \State Wait for last chunk to finish assembly
  \EndParFor
  \State \Return matrix $A$
  \EndProcedure
\end{algorithmic}
\vspace{-8pt}
\hrulefill
\end{algorithm}

\subsubsection{Global matrix assembly}
During the global assembly procedure, the raw data are finally merged, i.e. the contributions from individual boundary elements are added to the global degrees of freedom. Since multiple parallel processes may attempt to add their integral values to the same matrix entry at the same time, the access to the corresponding memory locations needs to be controlled via \textit{mutexes}. With the aim to minimize the inactive times when threads are waiting for their turn, every single matrix entry is protected by its own mutex variable. Alternative strategies employ only one mutex for the whole matrix, or column- and row-wise mutexes, respectively. However, our fine-grained approach has proven most effective. Note that this is not an issue with piecewise constant basis functions. In this case, matrix entries are assigned to exactly one pair of elements, thus parallel threads do not influence each other.

\subsubsection{Handling of different types of integrals}
To obtain sufficiently accurate evaluations of the discrete BEM kernels between test and trial elements suitable quadrature routines need to be chosen. If the test and trial element are disjoint a standard Gauss triangle quadrature rule is sufficient. For neighboring triangles (sharing a vertex or an edge) or in the case that the test and trial element are identical, suitable transformations need to be performed to remove the singularities. In our implementation we use a fully numerical scheme proposed by \citet{erichsen1998}. 

Only results related to regular integrals of disjoint elements are taken from the GPU computation. Otherwise, element pairs assigned to threads of a warp may choose different execution paths according to suitable integration routines, which leads to branch divergence on the device. This violates the SIMD paradigm of the GPU hardware architecture, and therefore harms performance. A possible remedy would be to presort the element pairs with respect to their integration scheme, and then call one CUDA kernel for each batch separately. However, the necessary preparations are not only cumbersome, but also fragment the work package resulting in multiple less efficient kernel calls. For this reason, we forgo singular integration on the device in this work. Instead, values related to singular integrals are simply overridden with cached results that have been evaluated on the CPU in advance.

\subsubsection{Data precision}
Due to the often significantly higher single precision performance on GPUs our implementation allows to switch the data type on the device between single and double precision. One argument is the restricted accuracy of the Gauss quadrature, which usually does not exceed single-precision accuracy.

\subsubsection{Treatment of complex numbers}
The last comment concerns the handling of complex-valued (Helmholtz) boundary integral operators. In this case, the real and imaginary part of complex numbers are consequently separated on the device. Although complex data types are available in CUDA, for instance as a part of the Thrust Parallel Algorithms Library \cite{thrustLibrary}, \citet{hawick2011numerical} have already shown that calculating and storing real and imaginary parts separately should be preferred over a compound data type for the sake of performance.

\subsection{GPU implementation of the numerical integration routine}
\label{subsec:denseImplGPU}

In the implementation of the GPU integration routine, the given task is split up into two subsequent steps, the evaluation of kernel function values and the summation procedure. These are processed within a single CUDA kernel call, where one thread is assigned to one pair of elements.

Since all element-related input data such as normal vectors, global quadrature points, and Jacobian determinants have been precalculated in advance, these information are simply loaded to thread-local memory. The BEM kernel function is then evaluated for all pairs of quadrature points, where the intermediate results are stored locally. Afterwards, the actual numerical integration procedure is triggered. The kernel function values are multiplied by quadrature weights, and basis functions values are incorporated for the different combinations of local degrees of freedom. Finally, the results are written back to global device memory in a coalesced fashion.
\begin{algorithm}[H]
\hrulefill
\vspace{-4pt}
\begin{algorithmic}
  \Procedure{Integral Evaluation~}{element pair $\left(m,n\right)$}
    \State Reconstruct test and trial element indices from thread index
    \State Load input data into thread-local device memory
    \ForAll{pairs of Gaussian points $\left(x_m,y_n\right)$}
      \State Evaluate kernel function $g\left(x_m,y_n\right)$
    \EndFor
    \ForAll{combinations of local dofs $\left(d_m,e_n\right)$}
      \State Evaluate integral $r\left(d_m,e_n\right)$
      \State Store results $b \gets r\left(d_m,e_n\right)$
    \EndFor
  \EndProcedure
\end{algorithmic}
\vspace{-8pt}
\hrulefill
\end{algorithm}
\section{GPU acceleration of the H-Matrix assembly}
\label{sec:hmatImpl}

Implementing ACA-based \H-Matrix compression efficiently on the GPU is not straight forward.
The iterative ACA scheme with per-block adaptive error control leads to bad load balancing and branch divergence issues inside a warp. A possible solution is to perform only the actual integral evaluation on the GPU, and keep the complex flow control on the CPU. However, this means sending only a single row or column of a block to the device at a time, leading to excessive kernel calls or underutilization. In \cite{boerm2015approximation} an approach is suggested in which a thread traverses the \H-Matrix tree and adds blocks to a task list. Once a task list is sufficiently large, another thread sends it to the GPU device for processing. While this approach is effective, it is implementationally difficult with significant bookkeeping and thread management and requires reorganization of the existing \H-Matrix code, which we aim to avoid here.

In our proposed GPU-accelerated \H-Matrix assembly each block is handled by one CPU thread. These parallel threads send integration jobs directly to the device without switching the context such that all data stays local to one CPU thread and any bookkeeping is omitted. Traditionally, the kernel calls on the GPU would have been serialized by the CUDA device and then executed one after another, leading again to underutilization problems as each thread only sends a row or column. The solution is to make use of NVIDIA's Hyper-Q technology. Hyper-Q was introduced with the Kepler generation of NVIDIA GPUs and increases the number of hardware work queues from one to thirty-two. This allows multiple CPU threads or processes to send smaller amounts of data to the device whereas the independent work queues ensure that device utilization remains high.

Moreover, inter-thread communication is avoided. This approach is comparatively simple to implement and it fits well with existing CPU-based \H-Matrix implementations which usually use threading to split up blocks between different CPU cores. The implementation of the row and column computation can simply build on the existing dense GPU integrator from Section \ref{sec:denseImpl}.

\subsection{Basic procedure}

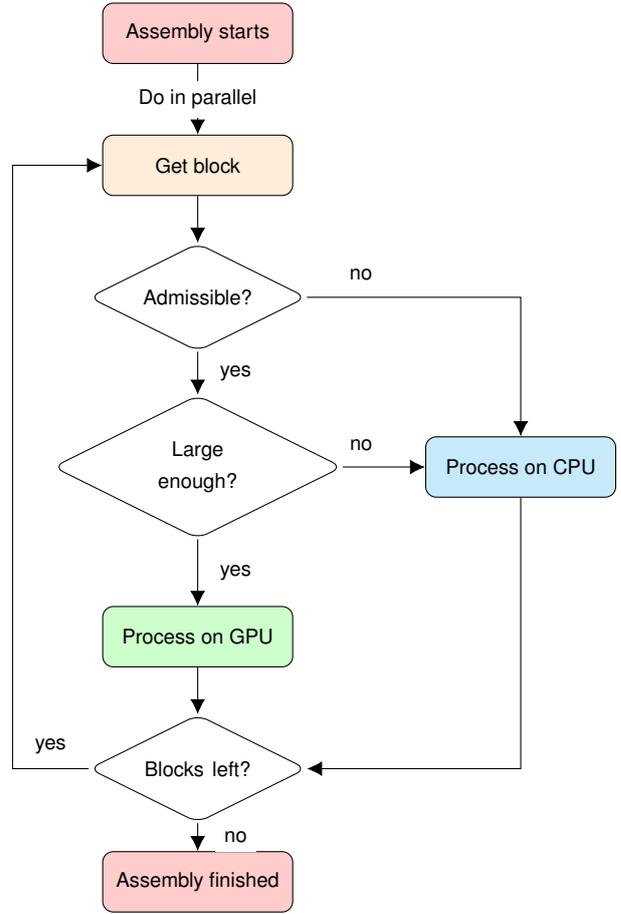
\begin{figure}[t]
  \centering
  \tikzset{%
  >={Latex[width=2mm,length=2mm]},
            line/.style = {draw, ->},
            base/.style = {rectangle, rounded corners, draw=black,
                           minimum width=2.5cm, minimum height=0.8cm,
                           text centered, font=\sffamily},
        hardware/.style = {rectangle, draw=black,
                           minimum width=0.8cm, minimum height=0.8cm,
                           text centered, font=\sffamily},
       startstop/.style = {base, fill=red!20},
       integrate/.style = {base, fill=green!20},
        assemble/.style = {base, fill=cyan!20},
         process/.style = {base, fill=orange!15},
              test/.style={base, diamond, aspect=2, text width=5em},
}

\begin{tikzpicture}[node distance=1.25cm,
    every node/.style={fill=white, font=\sffamily}, align=center]
    
  \node (start)             [startstop]     {\footnotesize Assembly starts};
  \node (getBlock)          [process, below of=start, yshift=-0.5cm]
                                                  {\footnotesize Get block};
  \node (admTest)        	[test, below of=getBlock, yshift=-0.5cm]
                                                {\footnotesize Admissible?};
  \node (sizeTest)          [test, below of=admTest, yshift=-1cm]
                                              {\footnotesize Large enough?};
  \node (gpuBlock)          [integrate, below of=sizeTest, yshift=-1cm]
                                              {\footnotesize Process on GPU};
  \node (cpuBlock)			[assemble, right of=sizeTest, xshift=3cm]
                                              {\footnotesize Process on CPU};
  \node (stopTest)          [test, below of=gpuBlock, yshift=-0.5cm]
                                                {\footnotesize Blocks left?};
  \node (ActivityDestroyed) [startstop, below of=start, yshift=-10cm]
                                           {\footnotesize Assembly finished};
                                           
  \draw[->]				(start) -- node [] {\footnotesize Do in parallel}
                                   (getBlock);
  \draw[->]          (getBlock) -- (admTest);    
  \draw[->]           (admTest) -- node [xshift=0.5cm] {\footnotesize yes}
                                   (sizeTest);
  \draw[->]           (admTest) -| node [xshift=-2.1cm, yshift=0.3cm]
                                   {\footnotesize no} (cpuBlock);
  \draw[->]          (sizeTest) -- node [xshift=0.5cm] {\footnotesize yes}
                                   (gpuBlock);
  \draw[->]          (sizeTest) -- node [yshift=0.3cm, xshift=-0.3cm]
                                   {\footnotesize no} (cpuBlock);
  \draw[->]          (gpuBlock) -- (stopTest);
  \draw[->]          (cpuBlock) |- (stopTest);
  \draw[->]          (stopTest) -- node [xshift=0.5cm]
                                   {\footnotesize no} (ActivityDestroyed);
  \draw[->]     (stopTest.west) -- node [yshift=0.3cm] {\footnotesize yes}
                                ++ (-1cm, 0) |- (getBlock);
\end{tikzpicture}
  \caption{\H-Matrix weak-form assembly procedure}
  \label{fig:hmatAssemblyProcedure}
\end{figure}

The GPU implementation of the \H-Matrix weak-form assembly comprises the following steps. First, the tree-based block partitioning of the \H-Matrix is generated on the CPU. This is a purely geometry-based computation and is very fast. Then, the participating GPUs are initialized. The grid data is copied to the device memory of the GPU, where element-related data is precalculated and cached in the global device memory. Afterwards, a thread-parallel loop over all \H-Matrix blocks is performed. Small non-admissible blocks involving singular integrals are treated in dense mode on the CPU. The remaining admissible blocks making up the largest part are fed to the ACA algorithm. For smaller blocks the ACA is completely performed on the CPU as the memory transfer overhead would be too large. The rows and columns of larger blocks are sent to the GPU during the course of the ACA assembly. The following pseudo code demonstrates this algorithm (see also Figure \ref{fig:hmatAssemblyProcedure}.
\begin{algorithm}[H]
\hrulefill
\vspace{-4pt}
\begin{algorithmic}
  \Procedure{H-Matrix Assembly}{}
  \State Declare H-Matrix $A$
  \State Generate block cluster tree
  \ParForAll{leaf blocks $b$}
    \If{$b$ is not admissible}
      \State Evaluate block in dense mode on CPU
    \ElsIf{$b$ is admissible}
      \State Approximate block using ACA
      \If{$b$ is large enough}
        \State Compute rows/columns of $b$ on GPU
      \ElsIf{$b$ is small}
        \State Compute rows/columns of $b$ on CPU
      \EndIf
    \EndIf
    \State H-Matrix $A \gets$ block $b$
  \EndParFor
  \State \Return H-Matrix $A$
\EndProcedure
\end{algorithmic}
\vspace{-8pt}
\hrulefill
\end{algorithm}
The advantage is that the code is very similar to an existing pure CPU thread-parallel ACA assembly routine. The difficult work of scheduling the rows and columns on the GPU is automatically performed through the hardware worker queues. Software management of the workload is not necessary.

If multiple GPUs are used we employ a simple round-robin strategy that passes the work packages around the participating GPUs. The local \H-Matrix assembler works in a very similar way to the dense assembler. First, the work package consisting of test-trial element pairs is split up into smaller chunks if their number exceeds a user-defined maximum. After that, memory is allocated both on the host and on the device to store the unassembled results. In contrast to the dense-matrix weak-form assembly, we use simple \textit{pageable} host memory and global device memory in this case. Therefore, results need to be copied to the host system explicitly, as we can not rely on the implicit way in conjunction with pinned host memory and zero-copy strategies. The reason for these different choices of host memory is that during the \H-Matrix construction memory is allocated and deallocated frequently, every time a row or column of a block needs to be computed. In the dense-matrix case this is done only once at the beginning and at the end of the whole procedure. Since allocation and deallocation of pinned host memory is quite expensive, it is way more efficient to stick with ordinary pageable host memory in this case.

Next, the test and trial element indices are copied to the global device memory, and a parallel task group is created to perform the numerical integration on the device simultaneously with the assembly of previously calculated values. A loop over all chunks finally constructs the requested part of a matrix block.

Note that the employed numerical integration routines are exactly the same as in the dense-matrix assembly case. The corresponding CUDA kernel has been described in detail in Section \ref{subsec:denseImplGPU}.

\subsection{A remark on compiler parameters}

It is important to add the \verb+-default-stream per-thread+ compiler flag to the \verb+nvcc+ command. This essentially enables the Hyper-Q technology and ensures that each host thread accessing a GPU creates its own default stream \cite{harrisStreamsConcurrency}, and thus multiple host-controlled calculations can be performed on one device concurrently.
\section{Realistic performance benchmarks}
\label{sec:benchmark}

\subsection{Hard- and software configuration}
\label{subsec:benchmarkHardware}

The benchmark tests are based on an experimental GPU enabled version of Bempp 3.1 (\url{www.bempp.com}). The code is run on a desktop workstation equipped with two Intel Xeon processors E5-2670 v2 \cite{intelXeonProcessorSpecs} operating at $\SI{2.5}{\giga\hertz}$ processor base frequency, as well as two NVIDIA GeForce GTX TITAN Black GPUs \cite{nvidiaTitanGPUSpecs} with a Kepler GK110 GPU each. We use the CUDA Toolkit \cite{cudaToolkit} version 8.0. The size of the system memory is about $\SI{200}{\giga\byte}$. We note that the aim of this section is to perform realistic benchmark tests.  Hence, we are not interested in comparing the GPU setup against a single CPU thread computation, but the overall performance of the GPU setup compared to a modern dual Xeon workstation that is typical for many BEM application scenarios.

\subsection{Experimental setup and procedure}
\label{subsec:benchmarkSetup}

We determine the performance gains over the multi-threaded CPU-optimized version of the code in terms of a speed-up factor
\begin{align}
  S = \frac{t_{\mathrm{CPU}}}{t_{\mathrm{GPU}}} \,,
  \label{eq:speedup}
\end{align}
where $t_{\mathrm{CPU}}$ is the execution time of the CPU and $t_{\mathrm{GPU}}$ is the execution time for the GPU enabled algorithms presented in the previous sections.

As benchmark problem we consider a unit sphere, which is discretized with various numbers of plane triangular boundary elements. Both the trial and the test space are restricted to globally continuous, piecewise linear basis functions. Note that basically all types of boundary elements provided by the Bempp library are supported on GPUs as a result of this work. However, other types of boundary elements do not introduce any significantly new aspects here. The number of unknowns $N$ arise from convenient mesh sizes $h$, where the smallest value of $h$ is chosen to exploit the available resources (i.e. the host memory) as much as possible. Our workstation features about $\SI{200}{\giga\byte}$ system memory. We then compute the discretized weak formulations of the three-dimensional Laplace and Helmholtz boundary operators as introduced in Section \ref{sec:bempp}, where both dense and \H-Matrices are employed. Each configuration is run 5 times to give averaged performance results.

\subsection{Dense-matrix weak-form assembly results}
\label{subsec:denseBenchmark}

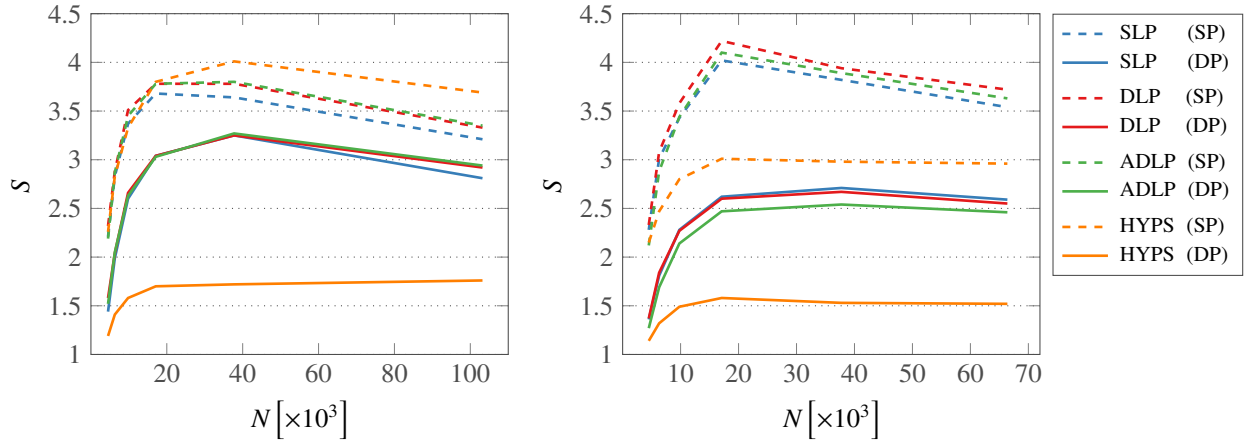
\begin{figure*}[t]
  \setlength\figurewidth{0.43\textwidth}
  \begin{subfigure}{0.42\textwidth}
    \begin{tikzpicture}


\begin{axis}[
width=\figurewidth,
every outer x axis line/.append style={white!30!black},
every x tick label/.append style={font=\color{white!30!black}},
xmin=0,
xmax=110,
xminorticks=true,
xlabel={$N\left[\times\num{e3}\right]$},
x unit=,
xtick={20,40,...,100},
every outer y axis line/.append style={white!30!black},
every y tick label/.append style={font=\color{white!30!black}},
ymin=1.0,
ymax=4.5,
ytick={1,1.5,...,4.5},
yminorticks=true,
ylabel={$S$},
ymajorgrids,
y unit=,
enlarge x limits=false,
enlarge y limits=false,
legend style={
  at={(0.5,-0.3)},
  anchor=north,
  legend cell align=left,
  align=left,
  fill=none,
  draw=none,
  legend columns=4,
  row sep=4pt,
  font=\footnotesize
},
grid style={dotted,white!30!black}
]

\addplot[color=niceblue,dashed,line width=1.0pt,mark size=5.0pt,mark=none,mark options={solid}]
  table[row sep=crcr]{%
4.51	2.21	\\
6.30	2.83	\\
9.79	3.37	\\
17.09	3.68	\\
37.74	3.64	\\
103.21	3.21	\\
};

\addplot[color=nicered,dashed,line width=1.0pt,mark size=5.0pt,mark=none,mark options={solid}]
  table[row sep=crcr]{%
4.51	2.32	\\
6.30	2.88	\\
9.79	3.51	\\
17.09	3.78	\\
37.74	3.78	\\
103.21	3.33	\\
};

\addplot[color=nicegreen,dashed,line width=1.0pt,mark size=5.0pt,mark=none,mark options={solid}]
  table[row sep=crcr]{%
4.51	2.19	\\
6.30	2.87	\\
9.79	3.45	\\
17.09	3.78	\\
37.74	3.80	\\
103.21	3.35	\\
};

\addplot[color=niceorange,dashed,line width=1.0pt,mark size=5.0pt,mark=none,mark options={solid}]
  table[row sep=crcr]{%
4.51	2.26	\\
6.30	2.83	\\
9.79	3.33	\\
17.09	3.80	\\
37.74	4.01	\\
103.21	3.69	\\
};

\addplot[color=niceblue,solid,line width=1.0pt,mark size=5.0pt,mark=none,mark options={solid}]
  table[row sep=crcr]{%
4.51	1.44	\\
6.30	1.99	\\
9.79	2.60	\\
17.09	3.04	\\
37.74	3.25	\\
103.21	2.81	\\
};

\addplot[color=nicered,solid,line width=1.0pt,mark size=5.0pt,mark=none,mark options={solid}]
  table[row sep=crcr]{%
4.51	1.58	\\
6.30	2.04	\\
9.79	2.66	\\
17.09	3.04	\\
37.74	3.25	\\
103.21	2.92	\\
};

\addplot[color=nicegreen,solid,line width=1.0pt,mark size=5.0pt,mark=none,mark options={solid}]
  table[row sep=crcr]{%
4.51	1.52	\\
6.30	2.05	\\
9.79	2.63	\\
17.09	3.03	\\
37.74	3.27	\\
103.21	2.94	\\
};

\addplot[color=niceorange,solid,line width=1.0pt,mark size=5.0pt,mark=none,mark options={solid}]
  table[row sep=crcr]{%
4.51	1.19	\\
6.30	1.41	\\
9.79	1.58	\\
17.09	1.70	\\
37.74	1.72	\\
103.21	1.76	\\
};

\end{axis}
\end{tikzpicture}%
  \end{subfigure}
  \begin{subfigure}{0.56\textwidth}
    \begin{tikzpicture}

\begin{axis}[
width=\figurewidth,
every outer x axis line/.append style={white!30!black},
every x tick label/.append style={font=\color{white!30!black}},
xmin=0,
xmax=72,
xminorticks=true,
xlabel={$N\left[\times\num{e3}\right]$},
x unit=,
xtick={10,20,...,70},
every outer y axis line/.append style={white!30!black},
every y tick label/.append style={font=\color{white!30!black}},
ymin=1.0,
ymax=4.5,
ytick={1,1.5,...,4.5},
yminorticks=true,
ylabel={$S$},
ymajorgrids,
y unit=,
enlarge x limits=false,
enlarge y limits=false,
legend style={
  at={(1.03,1.0)},
  anchor=north west,
  legend cell align=left,
  align=left,
  fill=none,
  draw=white!30!black,
  legend columns=1,
  font=\footnotesize,
  /tikz/row 2/.append style={yshift=8pt},
  /tikz/row 4/.append style={yshift=8pt},
  /tikz/row 6/.append style={yshift=8pt}
},
grid style={dotted,white!30!black}
]

\addplot[color=niceblue,dashed,line width=1.0pt,mark size=5.0pt,mark=none,mark options={solid}]
  table[row sep=crcr]{%
4.51	2.28	\\
6.30	3.01	\\
9.79	3.43	\\
17.09	4.02	\\
37.74	3.82	\\
66.38	3.54	\\
};
\addlegendentry{~SLP \hspace{9.5pt}(SP)};

\addplot[color=niceblue,solid,line width=1.0pt,mark size=5.0pt,mark=none,mark options={solid}]
  table[row sep=crcr]{%
4.51	1.36	\\
6.30	1.81	\\
9.79	2.28	\\
17.09	2.62	\\
37.74	2.71	\\
66.38	2.59	\\
};
\addlegendentry{~SLP \hspace{9.5pt}(DP)};

\addplot[color=nicered,dashed,line width=1.0pt,mark size=5.0pt,mark=none,mark options={solid}]
  table[row sep=crcr]{%
4.51	2.33	\\
6.30	3.08	\\
9.79	3.58	\\
17.09	4.22	\\
37.74	3.94	\\
66.38	3.72	\\
};
\addlegendentry{~DLP \hspace{7.5pt}(SP)};

\addplot[color=nicered,solid,line width=1.0pt,mark size=5.0pt,mark=none,mark options={solid}]
  table[row sep=crcr]{%
4.51	1.37	\\
6.30	1.84	\\
9.79	2.27	\\
17.09	2.60	\\
37.74	2.67	\\
66.38	2.55	\\
};
\addlegendentry{~DLP \hspace{7.5pt}(DP)};

\addplot[color=nicegreen,dashed,line width=1.0pt,mark size=5.0pt,mark=none,mark options={solid}]
  table[row sep=crcr]{%
4.51	2.12	\\
6.30	2.88	\\
9.79	3.44	\\
17.09	4.10	\\
37.74	3.89	\\
66.38	3.63	\\
};
\addlegendentry{~ADLP \hspace{2.5pt}(SP)};

\addplot[color=nicegreen,solid,line width=1.0pt,mark size=5.0pt,mark=none,mark options={solid}]
  table[row sep=crcr]{%
4.51	1.27	\\
6.30	1.69	\\
9.79	2.14	\\
17.09	2.47	\\
37.74	2.54	\\
66.38	2.46	\\
};
\addlegendentry{~ADLP \hspace{2.5pt}(DP)};

\addplot[color=niceorange,dashed,line width=1.0pt,mark size=5.0pt,mark=none,mark options={solid}]
  table[row sep=crcr]{%
4.51	2.15	\\
6.30	2.47	\\
9.79	2.80	\\
17.09	3.01	\\
37.74	2.98	\\
66.38	2.96	\\
};
\addlegendentry{~HYPS \hspace{2.5pt}(SP)};

\addplot[color=niceorange,solid,line width=1.0pt,mark size=5.0pt,mark=none,mark options={solid}]
  table[row sep=crcr]{%
4.51	1.14	\\
6.30	1.32	\\
9.79	1.49	\\
17.09	1.58	\\
37.74	1.53	\\
66.38	1.52	\\
};
\addlegendentry{~HYPS \hspace{2.5pt}(DP)};

\end{axis}
\end{tikzpicture}%
  \end{subfigure}
  \caption{Speed-up values $S$ for the dense weak-form assembly of Laplace (left) and Helmholtz (right) boundary operators depending on the number of unknowns $N$ in single precision (dashed lines) and double precision (solid lines)}
  \label{fig:denseWeakFormAssemblySpeedups}
\end{figure*}
Figure \ref{fig:denseWeakFormAssemblySpeedups} summarizes the computational results for the dense-matrix assembly of boundary operators. Initially, the speed-up rises steeply as the problem size grows, since any fixed overhead arising from the involvement of GPUs is spread over a larger number of unknowns. Then, it gradually levels off, and even shows a slight decrease in most cases. This can be ascribed to a reduced GPU clock frequency due to an increased mean GPU temperature under load. As heat is known to adversely affect performance when power consumption is held constant \cite{price2014temperature}, NVIDIAs GPU Boost 2.0 technology \cite{nvidiaGPUboost} applies dynamic overclocking with a temperature target of $\SI{80}{\celsius}$. This feature can not be disabled for GeForce GTX TITAN graphics cards operated under Debian Linux to enforce a constant clock frequency. The speed-up is likely to go up again when the GPUs eventually operate at their base frequency. However, due to memory limitations this can not be shown here.

As expected, GPU computations in single precision perform significantly faster than in double precision. This is particularly pronounced for complex-valued (Helmholtz) boundary operators. Our graphics boards provide three times as many single-precision floating-point units per multiprocessor as double-precision cores, which partly translates into the performance results. Issues of memory access and data transfer do not allow for an equivalent speed-up factor, though.

In comparison with the SLP, DLP and ADLP boundary operators, the HYPS boundary operator shows considerably less satisfactory results. Unlike the other three operators, the integral evaluation associated with the hypersingular operator requires the computation of an inner vector product at the lowest level of the nested quadrature loops. Our kernel codes are not well adapted to handle the corresponding memory access pattern, since we have not optimized this further. The reason is that we can exploit the weak-form shown in \eqref{eq:hypsHelmholtz} to write the HYPS operator in the form
\begin{equation}
    \label{eq:hyp_via_slp}
    D = \sum_{j=1}^3 Q_j^T\hat{S}Q_j - k^2\sum_{j=1}^3 P_j^T\hat{S}P_j,
\end{equation}
where $\hat{S}$ is the representation of the single-layer potential operator in a space of discontinuous, element-wise linear basis functions, the matrices $Q_j$ are sparse matrices that represent the $j$-th curl-component of the trial functions, and the $P_j$ are sparse matrices that correspond to the product of the basis functions with the $j$-th component of the normal vector. While this formulation is not interesting for dense assembly, it provides benefits for the more application relevant \H-Matrix assembly since \H-Matrix assembly of the single-layer operator on a discontinuous space is in pure CPU experiments already about twice as fast as the assembly of the matrix $D$. However, the price is that the memory requirements grow by about a factor six due to the increased overall matrix size. More details to this approach and numerical experiments can be found in \cite{Betcke2017}.

\subsection{H-Matrix weak-form assembly results}
\label{subsec:hmatBenchmark}

\begin{figure*}[t]
  \setlength\figurewidth{0.43\textwidth}
  \begin{subfigure}{0.42\textwidth}
    \begin{tikzpicture}

\begin{axis}[
width=\figurewidth,
every outer x axis line/.append style={white!30!black},
every x tick label/.append style={font=\color{white!30!black}},
xmin=200,
xmax=4300,
xminorticks=true,
xlabel={$N\left[\times\num{e6}\right]$},
x unit=,
xtick={500,1000,...,4000},
xticklabels={$\num{0.5}$, $\num{1}$, $\num{1.5}$, $\num{2}$, $\num{2.5}$, $\num{3}$, $\num{3.5}$, $\num{4}$},
every outer y axis line/.append style={white!30!black},
every y tick label/.append style={font=\color{white!30!black}},
ymin=0.7,
ymax=5.8,
ytick={1,1.5,2,...,5.5},
yminorticks=true,
ylabel={$S$},
ymajorgrids,
y unit=,
enlarge x limits=false,
enlarge y limits=false,
grid style={dotted,white!30!black}
]

\addplot[color=niceblue,dashed,line width=1.0pt,mark size=5.0pt,mark=none,mark options={solid}]
  table[row sep=crcr]{%
303.25	1.66	\\
411.97	2.10	\\
593.01	2.21	\\
922.86	3.03	\\
1640.88	3.77	\\
4079.66	5.05	\\
};

\addplot[color=niceblue,solid,line width=1.0pt,mark size=5.0pt,mark=none,mark options={solid}]
  table[row sep=crcr]{%
303.25	1.50	\\
411.97	1.74	\\
593.01	1.85	\\
922.86	2.45	\\
1640.88	3.02	\\
4079.66	3.77	\\
};

\addplot[color=nicered,dashed,line width=1.0pt,mark size=5.0pt,mark=none,mark options={solid}]
  table[row sep=crcr]{%
303.25	1.71	\\
411.97	2.17	\\
593.01	2.29	\\
922.86	2.92	\\
1640.88	3.99	\\
4079.66	5.46	\\
};

\addplot[color=nicered,solid,line width=1.0pt,mark size=5.0pt,mark=none,mark options={solid}]
  table[row sep=crcr]{%
303.25	1.51	\\
411.97	1.91	\\
593.01	1.93	\\
922.86	2.44	\\
1640.88	3.25	\\
4079.66	4.25	\\
};

\addplot[color=nicegreen,dashed,line width=1.0pt,mark size=5.0pt,mark=none,mark options={solid}]
  table[row sep=crcr]{%
303.25	1.64	\\
411.97	2.15	\\
593.01	2.21	\\
922.86	3.05	\\
1640.88	3.74	\\
4079.66	4.82	\\
};

\addplot[color=nicegreen,solid,line width=1.0pt,mark size=5.0pt,mark=none,mark options={solid}]
  table[row sep=crcr]{%
303.25	1.46	\\
411.97	1.81	\\
593.01	1.92	\\
922.86	2.51	\\
1640.88	3.03	\\
4079.66	3.81	\\
};

\addplot[color=niceorange,dashed,line width=1.0pt,mark size=5.0pt,mark=none,mark options={solid}]
  table[row sep=crcr]{%
303.25	1.22	\\
411.97	1.44	\\
593.01	1.07	\\
922.86	1.13	\\
1640.88	1.35	\\
};

\addplot[color=niceorange,solid,line width=1.0pt,mark size=5.0pt,mark=none,mark options={solid}]
  table[row sep=crcr]{%
303.25	1.19	\\
411.97	1.45	\\
593.01	1.35	\\
922.86	1.56	\\
1640.88	1.99	\\
};

\end{axis}
\end{tikzpicture}%
  \end{subfigure}
  \begin{subfigure}{0.56\textwidth}
    \begin{tikzpicture}

\begin{axis}[
width=\figurewidth,
every outer x axis line/.append style={white!30!black},
every x tick label/.append style={font=\color{white!30!black}},
xmin=200,
xmax=3300,
xminorticks=true,
xlabel={$N\left[\times\num{e6}\right]$},
x unit=,
xtick={500,1000,...,3000},
xticklabels={$\num{0.5}$, $\num{1}$, $\num{1.5}$, $\num{2}$, $\num{2.5}$, $\num{3}$},
every outer y axis line/.append style={white!30!black},
every y tick label/.append style={font=\color{white!30!black}},
ymin=0.7,
ymax=5.8,
ytick={1,1.5,2,...,5.5},
yminorticks=true,
ylabel={$S$},
ymajorgrids,
y unit=,
enlarge x limits=false,
enlarge y limits=false,
legend style={
  at={(1.03,1.0)},
  anchor=north west,
  legend cell align=left,
  align=left,
  fill=none,
  draw=white!30!black,
  legend columns=1,
  font=\footnotesize,
  /tikz/row 2/.append style={yshift=8pt},
  /tikz/row 4/.append style={yshift=8pt},
  /tikz/row 6/.append style={yshift=8pt}
},
grid style={dotted,white!30!black}
]

\addplot[color=niceblue,dashed,line width=1.0pt,mark size=5.0pt,mark=none,mark options={solid}]
  table[row sep=crcr]{%
303.25	1.79	\\
411.97	2.13	\\
593.01	2.45	\\
922.86	2.99	\\
1640.88	3.88	\\
3044.17	5.08	\\
};
\addlegendentry{~SLP \hspace{9.5pt}(SP)};

\addplot[color=niceblue,solid,line width=1.0pt,mark size=5.0pt,mark=none,mark options={solid}]
  table[row sep=crcr]{%
303.25	1.51	\\
411.97	1.66	\\
593.01	1.83	\\
922.86	2.24	\\
1640.88	2.77	\\
3044.17	3.67	\\
};
\addlegendentry{~SLP \hspace{9.5pt}(DP)};

\addplot[color=nicered,dashed,line width=1.0pt,mark size=5.0pt,mark=none,mark options={solid}]
  table[row sep=crcr]{%
303.25	1.78	\\
411.97	2.08	\\
593.01	2.33	\\
922.86	3.03	\\
1640.88 4.01	\\
3044.17	4.35	\\
};
\addlegendentry{~DLP \hspace{7.5pt}(SP)};

\addplot[color=nicered,solid,line width=1.0pt,mark size=5.0pt,mark=none,mark options={solid}]
  table[row sep=crcr]{%
303.25	1.41	\\
411.97	1.66	\\
593.01	1.88	\\
922.86	2.33	\\
1640.88 3.02	\\
3044.17	3.67	\\
};
\addlegendentry{~DLP \hspace{7.5pt}(DP)};

\addplot[color=nicegreen,dashed,line width=1.0pt,mark size=5.0pt,mark=none,mark options={solid}]
  table[row sep=crcr]{%
303.25	1.73	\\
411.97	2.17	\\
593.01	2.31	\\
922.86	2.97	\\
1640.88 3.97	\\
3044.17	5.00	\\
};
\addlegendentry{~ADLP \hspace{2.5pt}(SP)};

\addplot[color=nicegreen,solid,line width=1.0pt,mark size=5.0pt,mark=none,mark options={solid}]
  table[row sep=crcr]{%
303.25	1.43	\\
411.97	1.78	\\
593.01	1.92	\\
922.86	2.38	\\
1640.88 2.99	\\
3044.17	3.48	\\
};
\addlegendentry{~ADLP \hspace{2.5pt}(DP)};

\addplot[color=niceorange,dashed,line width=1.0pt,mark size=5.0pt,mark=none,mark options={solid}]
  table[row sep=crcr]{%
303.25	1.52	\\
411.97	1.65	\\
593.01	1.78	\\
922.86	1.88	\\
1640.88	1.52	\\
};
\addlegendentry{~HYPS \hspace{2.5pt}(SP)};

\addplot[color=niceorange,solid,line width=1.0pt,mark size=5.0pt,mark=none,mark options={solid}]
  table[row sep=crcr]{%
303.25	1.33	\\
411.97	1.45	\\
593.01	1.51	\\
922.86	1.79	\\
1640.88	2.19	\\
};
\addlegendentry{~HYPS \hspace{2.5pt}(DP)};

\end{axis}
\end{tikzpicture}%
  \end{subfigure}
  \caption{Speed-up values $S$ for the \H-Matrix weak-form assembly of Laplace (left) and Helmholtz (right) boundary operators depending on the number of unknowns $N$ in single precision (dashed lines) and double precision (solid lines)}
  \label{fig:hmatWeakFormAssemblySpeedups}
\end{figure*}
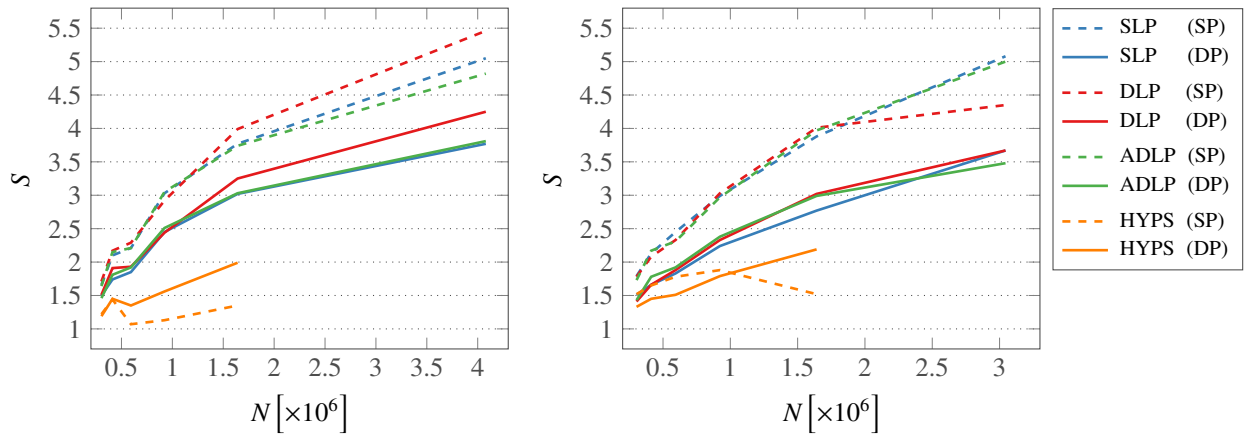

In the first part of this section we have focused on the assembly of the discrete weak form represented by a fully populated matrix. We now turn over to the fast approximation of boundary operators based on \H-Matrices. As in the previous section, both the Laplace and the Helmholtz equation are considered. The results are depicted in Figure \ref{fig:hmatWeakFormAssemblySpeedups}.

The kink in the line graphs substantiate the assumptions from the dense-assembly study that the slight decline in the speed-ups in Figure \ref{fig:denseWeakFormAssemblySpeedups} is a local phenomenon, rather than a global trend for the GPU-accelerated dense-matrix computation. The problem sizes, for which the mentioned significant changes in the underlying conditions occur (GPU clock rate), happen to coincide with the memory limit for dense-matrix weak-form representations. Here, the local depression is overcome, as much larger problems can be handled by means of the \H-Matrix approach.

Although the deviating behavior of the hypersingular operator has already been addressed in the course of the first part of this section, it is very striking here that the speed-ups related to single precision GPU computations tend to be lower than the factors obtained in the double precision case. This effect becomes stronger for larger problem sizes. An explanation can be found with a view to the number of iterations in the ACA algorithm. It turns out that, unlike the other operators, the HYPS operator is strongly affected by the precision level of the numerical integration on the GPU in the sense, that the approximation procedure for the admissible leaf blocks converges more slowly. Therefore, the total amount of work increases as more original matrix entries need to be computed. This can not be compensated by a faster numerical integration process in single precision, thus the \H-Matrix construction slows down as a whole compared to double precision computation. As a consequence, low precision numerical integration should be generally avoided in conjunction with hypersingular boundary operators where the weak form is based on \H-Matrices.

Note that the maximum speed-up that could be observed in Figure \ref{fig:hmatWeakFormAssemblySpeedups} is limited by the number of unknowns $N$, and therefore the size of the system memory in this case. However, the shape of the curves suggest that way higher factors can be realized for large-scale problems, provided that enough memory is available to store the \H-Matrix weak form.
\section{High-frequency scattering of a submarine hull}
\label{sec:application}

In order to demonstrate the benefits of our newly developed GPU-accelerated \H-Matrix BEM code to complex real-world problems we consider an exterior Helmholtz problem. In particular, we investigate the high-frequency scattered field from a plane wave impinging on a sound-hard submarine hull. To this end, a special OSRC-preconditioned Burton-Miller formulation \cite{antoine2007generalized, darbas2013, betcke2017_2} is applied to compute the \textit{bistatic target strength} of the submarine. This quantity indicates how detectable an object is with sonar in some distance, where the angular position of the transmitter and the receiver of the signal may differ. The ability to simulate the target strength characteristics is a prerequisite to develop construction features that give the submarine a desired low or high reflection coefficient already at an early stage in the design process.

The incident plane wave can be modeled as
\begin{align}
	u^{inc}\left(x\right) = u_{0}\exp\left(\mathrm{i}k\langle d,x\rangle\right) \,,
\end{align}
where $u_{0}$ is the amplitude, $d$ is the direction vector, $k$ denotes the wave number, and $x$ specifies a point in space.

The OSRC Burton-Miller formulation to compute the missing Dirichlet boundary data of the total field $\Phi = u^{\mathrm{sct}} + u^{\mathrm{inc}}$ reads
\begin{align}
	\left(\frac{1}{2}I-K-\tilde{V}_{NtD}D\right)\Phi = u^{inc} - \tilde{V}_{NtD}\frac{\partial u^{inc}}{\partial n_x} \,.
    \label{eq:osrcPrecondBurtonMillerBie}
\end{align}
Here, $\tilde{V}$ is an on-surface approximation to the Neumann-To-Dirichlet operator.
Further details on this OSRC preconditioning are given, for instance, by \citet{antoine2007generalized}.

Using the double-layer potential operator $K$, the scattered acoustic field simply yields
\begin{align}
	u^{\mathrm{sct}} = K\Phi \,.
\end{align}

We define a measure for the deviation of the GPU results from the CPU values based on $n$ evaluation points in the far field as
\begin{align}
	\Delta_{\mathrm{sct}} = \frac{1}{n} \sum_{i=1}^{n} \frac{\left|\left|u_{\mathrm{GPU}}\right|-\left|u_{\mathrm{CPU}}\right|\right|}{\left|u_{\mathrm{CPU}}\right|} \,.
    \label{eq:deviation}
\end{align}

Finally, the bistatic target strength is defined as
\begin{align}
	\mathrm{TS}_{\mathrm{dB}} = 20\log_{10}\left(R\left|\frac{u^{\mathrm{sct}}}{u_0}\right|\right) \,,
    \label{eq:ts}
\end{align}
with distance $R$ from the geometrical center of the investigated object.

In our example, we set the frequency of the incoming wave to $f=\SI{1000}{\hertz}$ and apply $c_{\mathrm{water}}=\SI{1500}{\meter\per\second}$, such that the wave number $k\approx 4.2$. Assuming that $\num{6}$ to $\num{10}$ boundary elements per wavelength $\lambda = \frac{c}{f}$ is enough to ensure good accuracy, we obtain a mesh with $M=\num{230000}$ linear plane triangles and $N=\num{115000}$ unknowns. This happens to be the largest problem that can be treated on our workstation in terms of system memory. The incident plane wave of amplitude $u_{0} = \SI{1}{\newton\per\square\meter}$ strikes the hull laterally at a $\theta^{\mathrm{inc}}=\SI{10}{\degree}$ angle from the submarine's longitudinal axis, thus the direction vector is defined as $d = \left[\cos\left(\theta^{\mathrm{inc}}\right), \sin\left(\theta^{\mathrm{inc}}\right), 0\right]^T$. Further, we position $\num{3600}$ evaluation points at a distance of $r=\SI{20}{\kilo\metre}$ from the geometrical center of the submarine model.
\begin{figure}[H]
  \centering
  \setlength\figurewidth{0.45\textwidth}
  \begin{tikzpicture}[
	scale=0.1,
    wavy/.style={decorate, decoration={snake, post length=1mm}},
    >={Latex[width=1.5mm,length=2mm]}
  ]
  
  \draw[fill=gray!30, draw=none] (0.0, 0.0) rectangle (50.0, 8.0);
  \draw (0.0, 0.0) -- (50.0, 0.0) (0.0, 8.0) -- (50.0, 8.0);
  \draw[fill=gray!30, draw=none] (0.0, 4.0) ellipse (6.0 and 4.0);
  \draw (0.0, 8.0) arc (90:270:6.0 and 4.0);
  \draw (8.0, 5.5) -- (8.5, 8.0) -- (11.0, 8.0) -- (11.0, 5.5);
  \draw (8.0, 2.5) -- (8.5, 0.0) -- (11.0, 0.0) -- (11.0, 2.5);
  \draw (25.0, 5.25) arc (65:295:5.0 and 1.25);
  \draw (25.0, 5.25) -- (33.0, 4.0) -- (25.0, 2.75);
  \draw[fill=gray!30] (55.0, 0.0) rectangle (58.0, 8.0);
  \filldraw[fill=gray!30, draw=none] (50.0, 8.0) -- (60.0, 4.0) -- (50.0, 0.0) -- cycle;
  \draw (50.0, 8.0) -- (60.0, 4.0) -- (50.0, 0.0);
  
  \draw[draw=gray] (-10.0, 4.0) arc (180:360:37.0);
  \draw[line width=3.5pt, line cap=round, dash pattern=on 0pt off 3.49\pgflinewidth]
  (-10.0, 4.0) arc (180:360:37.0);
  
  \draw[->, wavy] (-3.0 + 0 * 5 - 0.866 * 10.0, -2.0 - 0.5 * 10.0) -- (-3.0 + 0 * 5, -2.0);
  \draw[->, wavy] (-3.0 + 1 * 5 - 0.866 * 15.0, -2.0 - 0.5 * 15.0) -- (-3.0 + 1 * 5, -2.0);
  \draw[->, wavy] (-3.0 + 2 * 5 - 0.866 * 20.0, -2.0 - 0.5 * 20.0) -- (-3.0 + 2 * 5, -2.0);
  \draw[->, wavy] (-3.0 + 3 * 5 - 0.866 * 24.0, -2.0 - 0.5 * 25.0) -- (-3.0 + 3 * 5, -2.0);
  
  \draw[draw=black] (27.0, 6.0) -- (27.0, 2.0);
  \draw[draw=black] (25.0, 4.0) -- (29.0, 4.0);
  \draw[draw=gray] (27.0, 4.0) -- (-7.64, -16.0);
  \draw[draw=gray, ->] (27.0, 4.0) -- (59.0, -14.5);
  \node[] at (46.0, -15.0) {$r=\SI{20}{\kilo\metre}$};
  \draw[draw=gray] (-13.0, 4.0) -- (67.0, 4.0);
  \draw[draw=gray, ->] (14.0, 4.0) arc (180:210:13.0);
  \node[] at (15.0, -10.0) {$\theta^{\mathrm{inc}}=\SI{10}{\degree}$};
  \draw[draw=gray, ->] (17.0, 4.0) arc (180:330:10.0);
  \node[] at (36.0, -7.0) {$\theta^{\mathrm{sct}}$};
  
\end{tikzpicture}
  \caption{Experimental setup of the scattering submarine hull}
  \label{fig:submarineSetup}
\end{figure}
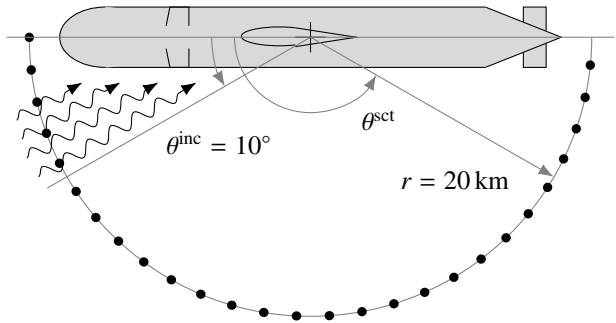
The threshold for rows or columns of \H-Matrix blocks to be treated by graphics processors is set to $\num{10000}$ participating element pairs. All GPU computations are performed in single precision.
The assembly of the hypersingular operator $D$ was implemented as demonstrated in \eqref{eq:hyp_via_slp} through sparse transformations of the single-layer boundary operator on a space of discontinuous linear functions. Direct assembly of the HYPS operator did not yield sufficient results in single-precision.
\begin{figure}[H]
  \centering
  \setlength\figurewidth{0.45\textwidth}
  \input{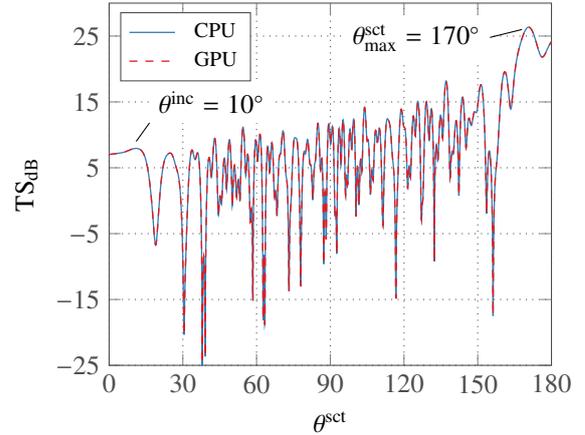}
  \caption{Bistatic target strength $\mathrm{TS}$ depending on the angular position $\theta^{\mathrm{sct}}$ at a distance $r=\SI{20}{\kilo\metre}$ from the submarine's center}
  \label{fig:submarineTsPlot}
\end{figure}
Figure \ref{fig:submarineTsPlot} shows the target strength of the submarine.
\begin{figure}[H]
  \centering
  \setlength\figurewidth{0.45\textwidth}
  \input{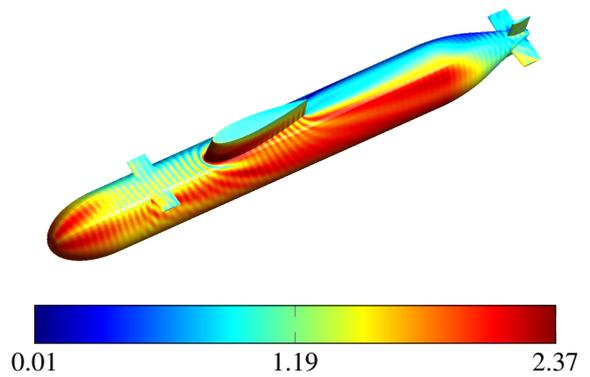}
  \caption{Dirichlet boundary data of the total field $\left|\Phi\right|$}
  \label{fig:submarineTotalFieldDirichletData}
\end{figure}
Figure \ref{fig:submarineTotalFieldDirichletData} visualizes the absolute values of the Dirichlet boundary data that has been solved for in equation \eqref{eq:osrcPrecondBurtonMillerBie}.

Using the GPU-accelerated assembly routines, the setup of the equation system took only $\SI{1206}{\second}$, compared to $\SI{2275}{\second}$ with the multi-threaded CPU-optimized code. This translates into a speed-up factor of $\num{1.9}$ for the discrete weak-form assembly of the required boundary operators, while the deviation between GPU and CPU results as defined by equation \eqref{eq:deviation} yields a negligible value of $\Delta_{\mathrm{sct}} = \SI{0.0081}{\percent}$. It is interesting to note how the assembly speed-up translates into a speed-up for the total application. The problem was solved in $\num{18}$ iterations due to the effectiveness of the OSRC preconditioner in this application. The assembly of all sparse operators and the solution of the iterative system were performed purely on the CPU. The total time (excluding reading the grid) including assembly, iterative solver and evaluation of the bistatic target strength with the GPU enabled code was \SI{2473}{\second} while the time for the pure CPU code was \SI{3397}{\second}, a total speed-up of a factor of $1.4$. Hence, the more the assembly dominates the total solution time the more will an application benefit from the simple GPU strategy described in this paper.
\section{Conclusions}
\label{sec:conclusions}
The aim of this paper was to develop a simple strategy to benefit from accelerated GPU computations for existing BEM codes without requiring a substantial rewrite. Indeed, only the core integration routines need to be replaced by corresponding GPU routines while the rest of the \H-Matrix assembly and surrounding code remains mostly unchanged. This allows an easy upgrade path for existing BEM codes to take advantage of GPU compute capabilities.

The performance advantage is considerable. A setup of gaming level dual Kepler generation boards outperformed a pure 20-core Xeon CPU workstation by a factor of $\num{1.9}$. This means that with modest investments in hardware and software development existing BEM codes can achieve significant speed-ups for a wide range of applications.

We note that while other authors have suggested GPU-accelerated \H-Matrix compression (see e.g. \cite{boerm2015approximation}), the approach taken here achieves good performance improvements without major rewrites by simply replacing the core assembly routines. The key difference is the availability of NVIDIA's Hyper-Q technology which simplifies code design for GPUs significantly. Moreover, we have provided a realistic application study that demonstrates how the assembly speed-up translates over from synthetic benchmarks to realistic application problems that additionally involve sparse operators and preconditioners, which are assembled on the CPU.

Further acceleration can be achieved by translating the complete tool chain including sparse operators and iterative solvers to the GPU. However, this requires substantial investments in redesigning the software infrastructure while at the same time classical CPU workstations are also increasing their speed considerably with modern high-end dual Xeon workstations providing over 50 physical CPU cores. We believe that the proposed algorithm in this paper provides a good balance between modest investments in hardware and software development while still achieving very good assembly speed-ups that translate into a noticeable overall application performance improvement.
\section*{Acknowledgement}

This work was supported by the Mobility Fund of Hamburg University of Technology.





\bibliographystyle{model1-num-names}
\bibliography{references.bib}

\begin{thebibliography}{30}
\expandafter\ifx\csname natexlab\endcsname\relax\def\natexlab#1{#1}\fi
\providecommand{\bibinfo}[2]{#2}
\ifx\xfnm\relax \def\xfnm[#1]{\unskip,\space#1}\fi
\bibitem[{Greengard and Rokhlin(1987)}]{greengard1987fast}
\bibinfo{author}{L.~Greengard}, \bibinfo{author}{V.~Rokhlin},
\newblock \bibinfo{title}{{A} {F}ast {A}lgorithm for {P}article {S}imulations},
\newblock \bibinfo{journal}{{J}ournal of {C}omputational {P}hysics}
  \bibinfo{volume}{73} (\bibinfo{year}{1987}) \bibinfo{pages}{325--348}.
\bibitem[{Liu(2009)}]{liu2009fast}
\bibinfo{author}{Y.~Liu}, \bibinfo{title}{{F}ast {M}ultipole {B}oundary
  {E}lement {M}ethod: {T}heory and {A}pplications in {E}ngineering},
  \bibinfo{publisher}{Cambridge University Press}, \bibinfo{year}{2009}.
\bibitem[{Keuchel et~al.(2017)Keuchel, Vater, and Estorff}]{keuchel2017hp}
\bibinfo{author}{S.~Keuchel}, \bibinfo{author}{K.~Vater},
  \bibinfo{author}{O.~v. Estorff},
\newblock \bibinfo{title}{hp {F}ast {M}ultipole {B}oundary {E}lement {M}ethod
  for 3{D} {A}coustics},
\newblock \bibinfo{journal}{{I}nternational {J}ournal for {N}umerical {M}ethods
  in {E}ngineering} \bibinfo{volume}{110} (\bibinfo{year}{2017})
  \bibinfo{pages}{842--861}.
\bibitem[{B{\"o}rm et~al.(2003)B{\"o}rm, Grasedyck, and
  Hackbusch}]{borm2003introduction}
\bibinfo{author}{S.~B{\"o}rm}, \bibinfo{author}{L.~Grasedyck},
  \bibinfo{author}{W.~Hackbusch},
\newblock \bibinfo{title}{{I}ntroduction to {H}ierarchical {M}atrices with
  {A}pplications},
\newblock \bibinfo{journal}{{E}ngineering {A}nalysis with {B}oundary
  {E}lements} \bibinfo{volume}{27} (\bibinfo{year}{2003})
  \bibinfo{pages}{405--422}.
\bibitem[{Hackbusch(1999)}]{hackbusch1999sparse}
\bibinfo{author}{W.~Hackbusch},
\newblock \bibinfo{title}{{A} {S}parse {Ma}trix {A}rithmetic {B}ased
  on$\backslash$ cal h-{M}atrices. {P}art {I}: {I}ntroduction to
  $\{$$\backslash$ Cal H$\}$-{M}atrices},
\newblock \bibinfo{journal}{{C}omputing} \bibinfo{volume}{62}
  (\bibinfo{year}{1999}) \bibinfo{pages}{89--108}.
\bibitem[{Takahashi and Hamada(2009)}]{takahashi2009gpu}
\bibinfo{author}{T.~Takahashi}, \bibinfo{author}{T.~Hamada},
\newblock \bibinfo{title}{{GPU}-{A}ccelerated {B}oundary {E}lement {M}ethod for
  {H}elmholtz' {E}quation in {T}hree {D}imensions},
\newblock \bibinfo{journal}{{I}nternational {J}ournal for {N}umerical {M}ethods
  in {E}ngineering} \bibinfo{volume}{80} (\bibinfo{year}{2009})
  \bibinfo{pages}{1295--1321}.
\bibitem[{Labaki et~al.(2011)Labaki, Ferreira, and
  Mesquita}]{labaki2011constant}
\bibinfo{author}{J.~Labaki}, \bibinfo{author}{L.~S. Ferreira},
  \bibinfo{author}{E.~Mesquita},
\newblock \bibinfo{title}{{C}onstant {B}oundary {E}lements on {G}raphics
  {H}ardware: {A} {GPU-CPU} {C}omplementary {I}mplementation},
\newblock \bibinfo{journal}{{J}ournal of the {B}razilian {S}ociety of
  {M}echanical {S}ciences and {E}ngineering} \bibinfo{volume}{33}
  (\bibinfo{year}{2011}) \bibinfo{pages}{475--482}.
\bibitem[{Labaki et~al.(2010)Labaki, Mesquita, and Ferreira}]{labaki2010bem}
\bibinfo{author}{J.~Labaki}, \bibinfo{author}{E.~Mesquita},
  \bibinfo{author}{S.~L. Ferreira},
\newblock \bibinfo{title}{{T}he {BEM} on {G}eneral {P}urpose {G}raphics
  {P}rocessing {U}nits ({GPGPU}): {A} {S}tudy on {T}hree {D}istinct
  {I}mplementations},
\newblock in: \bibinfo{booktitle}{Proceedings of the 11th International
  Conference on Boundary Element Techniques XI. United Kingdom: EC Ltd Press},
  pp. \bibinfo{pages}{316--323}.
\bibitem[{Yokota et~al.(2011)Yokota, Bardhan, Knepley, Barba, and
  Hamada}]{yokota2011biomolecular}
\bibinfo{author}{R.~Yokota}, \bibinfo{author}{J.~P. Bardhan},
  \bibinfo{author}{M.~G. Knepley}, \bibinfo{author}{L.~A. Barba},
  \bibinfo{author}{T.~Hamada},
\newblock \bibinfo{title}{{B}iomolecular {E}lectrostatics {U}sing a {F}ast
  {M}ultipole {BEM} on up to 512 {GPU}s and a {B}illion {U}nknowns},
\newblock \bibinfo{journal}{{C}omputer {P}hysics {C}ommunications}
  \bibinfo{volume}{182} (\bibinfo{year}{2011}) \bibinfo{pages}{1272--1283}.
\bibitem[{Yokota et~al.(2012)Yokota, Barba, Narumi, and
  Yasuoka}]{yokota2012scaling}
\bibinfo{author}{R.~Yokota}, \bibinfo{author}{L.~Barba},
  \bibinfo{author}{T.~Narumi}, \bibinfo{author}{K.~Yasuoka},
\newblock \bibinfo{title}{{S}caling {F}ast {M}ultipole {M}ethods up to 4000
  {GPU}s},
\newblock in: \bibinfo{booktitle}{Proceedings of the ATIP/A* CRC Workshop on
  Accelerator Technologies for High-Performance Computing: Does Asia Lead the
  Way?}, \bibinfo{organization}{A* STAR Computational Resource Centre},
  p.~\bibinfo{pages}{9}.
\bibitem[{B\"orm and Christophersen(2017)}]{boerm2015approximation}
\bibinfo{author}{S.~B\"orm}, \bibinfo{author}{S.~Christophersen},
\newblock \bibinfo{title}{{A}pproximation of {BEM} {U}sing {GPGPU}s},
\newblock \bibinfo{journal}{{CoRR}} \bibinfo{volume}{abs/1510.07244}
  (\bibinfo{year}{2017}).
\bibitem[{NVIDIA(2017)}]{cudaToolkit}
\bibinfo{author}{NVIDIA}, \bibinfo{title}{{CUDA} {T}oolkit},
  \bibinfo{howpublished}{\url{https://developer.nvidia.com/cuda-toolkit}},
  \bibinfo{year}{2017}. \bibinfo{note}{Accessed: 2017-11-02}.
\bibitem[{{\'S}migaj et~al.(2015){\'S}migaj, Betcke, Arridge, Phillips, and
  Schweiger}]{smigaj2015solving}
\bibinfo{author}{W.~{\'S}migaj}, \bibinfo{author}{T.~Betcke},
  \bibinfo{author}{S.~Arridge}, \bibinfo{author}{J.~Phillips},
  \bibinfo{author}{M.~Schweiger},
\newblock \bibinfo{title}{{S}olving {B}oundary {I}ntegral {P}roblems with
  {BEM}++},
\newblock \bibinfo{journal}{ACM Transactions on Mathematical Software (TOMS)}
  \bibinfo{volume}{41} (\bibinfo{year}{2015}) \bibinfo{pages}{6}.
\bibitem[{Intel(2017)}]{intelXeonProcessorSpecs}
\bibinfo{author}{Intel}, \bibinfo{title}{{I}ntel {X}eon {P}rocessor {E}5-2670
  v2 {S}pecifications},
  \bibinfo{howpublished}{\url{https://ark.intel.com/products/75275/Intel-Xeon-Processor-E5-2670-v2-25M-Cache-2_50-GHz}},
  \bibinfo{year}{2017}. \bibinfo{note}{Accessed: 2017-11-02}.
\bibitem[{NVIDIA(2017)}]{nvidiaTitanGPUSpecs}
\bibinfo{author}{NVIDIA}, \bibinfo{title}{{NVIDIA} {G}e{F}orce {GTX} {TITAN}
  {B}lack},
  \bibinfo{howpublished}{\url{http://www.geforce.com/hardware/desktop-gpus/geforce-gtx-titan-black}},
  \bibinfo{year}{2017}. \bibinfo{note}{Accessed: 2017-11-02}.
\bibitem[{Bebendorf(2008)}]{bebendorf2008hierarchical}
\bibinfo{author}{M.~Bebendorf}, \bibinfo{title}{{H}ierarchical {M}atrices},
  \bibinfo{publisher}{Springer}, \bibinfo{year}{2008}.
\bibitem[{NVIDIA(2017{\natexlab{a}})}]{cudaProgrammingGuide}
\bibinfo{author}{NVIDIA}, \bibinfo{title}{{CUDA} {C} {P}rogramming {G}uide},
  \bibinfo{howpublished}{\url{https://docs.nvidia.com/cuda/pdf/CUDA_C_Programming_Guide.pdf}},
  \bibinfo{year}{2017}{\natexlab{a}}. \bibinfo{note}{Accessed: 2017-11-02}.
\bibitem[{NVIDIA(2017{\natexlab{b}})}]{cudaBestPracticesGuide}
\bibinfo{author}{NVIDIA}, \bibinfo{title}{{CUDA} {C} {B}est {P}ractices
  {G}uide},
  \bibinfo{howpublished}{\url{https://docs.nvidia.com/cuda/pdf/CUDA_C_Best_Practices_Guide.pdf}},
  \bibinfo{year}{2017}{\natexlab{b}}. \bibinfo{note}{Accessed: 2017-11-02}.
\bibitem[{NVIDIA(2012)}]{nvidiaKeplerArch}
\bibinfo{author}{NVIDIA}, \bibinfo{title}{{NVIDIA}'s {N}ext {G}eneration {CUDA}
  {C}ompute {A}rchitecture: {K}epler {GK}110},
  \bibinfo{howpublished}{\url{https://www.nvidia.com/content/PDF/kepler/NVIDIA-Kepler-GK110-Architecture-Whitepaper.pdf}},
  \bibinfo{year}{2012}. \bibinfo{note}{Accessed: 2017-11-02}.
\bibitem[{Intel(2017)}]{intelTbb}
\bibinfo{author}{Intel}, \bibinfo{title}{{T}hreading {B}uilding {B}locks
  ({TBB})},
  \bibinfo{howpublished}{\url{https://www.threadingbuildingblocks.org/}},
  \bibinfo{year}{2017}. \bibinfo{note}{Accessed: 2017-11-02}.
\bibitem[{Erichsen and Sauter(1998)}]{erichsen1998}
\bibinfo{author}{S.~Erichsen}, \bibinfo{author}{S.~A. Sauter},
\newblock \bibinfo{title}{{E}fficient automatic quadrature in 3-d {G}alerkin
  {BEM}},
\newblock \bibinfo{journal}{{C}omputer {M}ethods in {A}pplied {M}echanics and
  {E}ngineering} \bibinfo{volume}{157} (\bibinfo{year}{1998})
  \bibinfo{pages}{215--224}. \bibinfo{note}{Papers presented at the Seventh
  Conference on Numerical Methods and Computational Mechanics in Science and
  Engineering}.
\bibitem[{Hoberock and Bell(2017)}]{thrustLibrary}
\bibinfo{author}{J.~Hoberock}, \bibinfo{author}{N.~Bell},
  \bibinfo{title}{{T}hrust {P}arallel {A}lgorithms {L}ibrary},
  \bibinfo{howpublished}{\url{https://thrust.github.io/}},
  \bibinfo{year}{2017}. \bibinfo{note}{Accessed: 2017-11-02}.
\bibitem[{Hawick and Playne(2011)}]{hawick2011numerical}
\bibinfo{author}{K.~A. Hawick}, \bibinfo{author}{D.~P. Playne},
\newblock \bibinfo{title}{{N}umerical {S}imulation of the {C}omplex
  {G}inzburg-{L}andau {E}quation on {GPU}s with {CUDA}},
\newblock in: \bibinfo{booktitle}{Proc. IASTED International Conference on
  Parallel and Distributed Computing and Networks (PDCN)}, pp.
  \bibinfo{pages}{39--45}.
\bibitem[{Harris(2015)}]{harrisStreamsConcurrency}
\bibinfo{author}{M.~Harris}, \bibinfo{title}{{GPU} {P}ro {T}ip: {CUDA} 7
  {S}treams {S}implify {C}oncurrency},
  \bibinfo{howpublished}{\url{https://devblogs.nvidia.com/parallelforall/gpu-pro-tip-cuda-7-streams-simplify-concurrency/}},
  \bibinfo{year}{2015}. \bibinfo{note}{Accessed: 2017-11-02}.
\bibitem[{Price(2014)}]{price2014temperature}
\bibinfo{author}{D.~Price}, \bibinfo{title}{{F}ire and {I}ce: {H}ow
  {T}emperature {A}ffects {GPU} {P}erformance},
  \bibinfo{howpublished}{\url{http://on-demand.gputechconf.com/gtc/2014/presentations/S4484-how-temperature-affects-gpu-performance.pdf}},
  \bibinfo{year}{2014}. \bibinfo{note}{GPU Technology Conference}.
\bibitem[{NVIDIA(2017)}]{nvidiaGPUboost}
\bibinfo{author}{NVIDIA}, \bibinfo{title}{{GPU} {B}oost 2.0 {T}echnology},
  \bibinfo{howpublished}{\url{http://www.nvidia.de/object/nvidia-gpu-boost-2-de.html}},
  \bibinfo{year}{2017}. \bibinfo{note}{Accessed: 2017-11-02}.
\bibitem[{Betcke et~al.(2017)Betcke, Scroggs, and \'Smigaj}]{Betcke2017}
\bibinfo{author}{T.~Betcke}, \bibinfo{author}{M.~W. Scroggs},
  \bibinfo{author}{W.~\'Smigaj},
\newblock \bibinfo{title}{{P}roduct algebras for {G}alerkin discretizations of
  boundary integral operators and their applications},
\newblock \bibinfo{journal}{in preparation}  (\bibinfo{year}{2017}).
\bibitem[{Antoine and Darbas(2007)}]{antoine2007generalized}
\bibinfo{author}{X.~Antoine}, \bibinfo{author}{M.~Darbas},
\newblock \bibinfo{title}{{G}eneralized {C}ombined {F}ield {I}ntegral
  {E}quations for the {I}terative {S}olution of the {T}hree-dimensional
  {H}elmholtz {E}quation},
\newblock \bibinfo{journal}{{ESAIM}: {M}athematical {M}odelling and {N}umerical
  {A}nalysis} \bibinfo{volume}{41} (\bibinfo{year}{2007})
  \bibinfo{pages}{147--167}.
\bibitem[{Darbas et~al.(2013)Darbas, Darrigrand, and Lafranche}]{darbas2013}
\bibinfo{author}{M.~Darbas}, \bibinfo{author}{E.~Darrigrand},
  \bibinfo{author}{Y.~Lafranche},
\newblock \bibinfo{title}{{C}ombining analytic preconditioner and {F}ast
  {M}ultipole {M}ethod for the 3-{D} {H}elmholtz equation},
\newblock \bibinfo{journal}{{J}ournal of {C}omputational {P}hysics}
  \bibinfo{volume}{236} (\bibinfo{year}{2013}) \bibinfo{pages}{289--316}.
\bibitem[{Betcke et~al.(2017)Betcke, {van 't Wout}, and
  G{\'{e}}lat}]{betcke2017_2}
\bibinfo{author}{T.~Betcke}, \bibinfo{author}{E.~{van 't Wout}},
  \bibinfo{author}{P.~G{\'{e}}lat},
\newblock \bibinfo{title}{{C}omputationally {E}fficient {B}oundary {E}lement
  {M}ethods for {H}igh-{F}requency {H}elmholtz {P}roblems in {U}nbounded
  {D}omains},
\newblock \bibinfo{publisher}{Birkh{\"{a}}user, Cham}, \bibinfo{year}{2017},
  pp. \bibinfo{pages}{215--243}.

\end{thebibliography}







\end{document}